\documentstyle[aaspp]{article}

\begin{document}

\def\MSUN{\rm M_{\odot}}
\def\RSUN{\rm R_{\odot}} 
\def\MSUNYR{\rm M_{\odot}\,yr^{-1}}
\def\MDOT{\dot{M}}

\newbox\grsign \setbox\grsign=\hbox{$>$} \newdimen\grdimen \grdimen=\ht\grsign
\newbox\simlessbox \newbox\simgreatbox
\setbox\simgreatbox=\hbox{\raise.5ex\hbox{$>$}\llap
     {\lower.5ex\hbox{$\sim$}}}\ht1=\grdimen\dp1=0pt
\setbox\simlessbox=\hbox{\raise.5ex\hbox{$<$}\llap
     {\lower.5ex\hbox{$\sim$}}}\ht2=\grdimen\dp2=0pt
\def\simgreat{\mathrel{\copy\simgreatbox}}
\def\simless{\mathrel{\copy\simlessbox}}

\title{ Dynamics of line-driven disk winds
in Active Galactic Nuclei.}

\vspace{1.cm}
\author{ Daniel Proga }
\vspace{.5cm}
\affil{LHEA, GSFC, NASA, Code 662, Greenbelt, MD 20771; proga@sobolev.gsfc.nasa.gov}

\author{ James M. Stone }
\vspace{.5cm}
\affil{Department of Astronomy, University of Maryland, College Park, MD 20742;
jstone@astro.umd.edu}

\author{Timothy R. Kallman  }
\vspace{.5cm}
\affil{LHEA, GSFC, NASA, Code 662, Greenbelt, MD 20771; tim@xstar.gsfc.nasa.gov}

\begin{abstract}
We present the results of axisymmetric time-dependent 
hydrodynamic calculations of line-driven winds from accretion disks
in  active galactic nuclei (AGN). We assume the disk is flat, Keplerian,
geometrically thin, and optically thick, radiating according to 
the $\alpha$-disk
prescription. 
The central engine of  the AGN is a source of both ionizing X-rays and
wind-driving ultraviolet (UV) photons. 
To calculate the radiation force, we take into account radiation
from the disk and the central engine. 
The gas temperature and ionization state in the wind are calculated 
self-consistently from the photoionization and heating rate of the central
engine.

We find that a disk accreting onto a $10^8~\MSUN$ black hole 
at the rate of $1.8~\MSUNYR$ can launch a 
wind at $\sim 10^{16}$~cm from the central engine. The X-rays 
from the central object are significantly attenuated by the disk
atmosphere so they 
cannot prevent the local disk radiation from pushing matter
away from the disk. 
However in the supersonic portion of the flow high above the disk, 
the X-rays can overionize the gas and decrease the wind terminal velocity.
For a reasonable X-ray opacity, 
e.g., $\kappa_{\rm X}=40~\rm g^{-1}~cm^2$, the disk 
wind can be accelerated by the central UV radiation to velocities of 
up to 15000~$\rm km~s^{-1}$ at a distance of $\sim 10^{17}$~cm 
from the central engine.
The covering factor of the disk wind is $\sim 0.2$. The wind is unsteady
and  consists of an opaque, slow vertical flow near 
the disk that is bounded on the polar side by a high-velocity stream. 
A typical column density 
through the fast stream is a few $10^{23}~\rm cm^{-2}$ so the stream 
is optically thin to the UV radiation.  This low column density
is precisely why gas can be accelerated
to high velocities. The fast stream contributes nearly 100\% to 
the total wind mass loss rate of $0.5~\MSUNYR$. 
\end{abstract}


\section{Introduction}

Many observed spectral features of active galactic nuclei (AGN) indicate
that  outflows are  common in these systems.
The prominent broad emission lines (BEL) in ultraviolet (UV) from
H~I, O~VI, N~V, C~IV, and Si~IV are the defining feature of  quasars   
(Blandford, Netzer \& Woltjer 1990;
Osterbrock 1989), and they may be associated with a high velocity outflow.
Fast outflows can also explain narrow UV absorption lines from highly 
ionized species
-- such as C~IV and N~V -- observed in approximately half of 
the HST-observed Seyfert 1 galaxies (e.g., Crenshaw 1997). 
These narrow absorption lines are  all blueshifted relative to the systemic
velocity by 0 to -1600~km~s$^{-1}$. Recent ASCA observations 
show that in a small sample observed by HST and ASCA, all of the Seyfert
galaxies with  warm absorbers also show intrinsic UV absorption.
A high-resolution X-ray observation of the Seyfert galaxy NGC~5548 
obtained by Chandra  shows strong, narrow absorption lines from highly
ionized species (Kaastra et al. 2000). 
The lines are blueshifted by a few hundred km~$\rm s^{-1}$
and they are reminiscent of the narrow absorption lines observed in the UV.
Some QSOs also  have  intrinsic UV narrow-line absorbers
with line-of-sight velocities as large as $~51000~{\rm km~s^{-1}}$
(Hamann et al. 1997).
Perhaps the most compelling evidence for fast outflows in QSOs
are strong broad absorption lines (BALs) in the UV
resonance lines of highly ionized species such as N~V, C~IV, and Si~IV.
The BALs are always blueshifted relative to the emission-line rest
frame, indicating the presence of outflows from the active nucleus,
with velocities as large as $60000~\rm km s^{-1}$.   

A key constraint on any model for the origin of AGN outflows is the 
ionization balance. On one hand we observe very high
luminosities in X-rays and the UV and 
on the other hand we observe spectral lines from moderately
and highly ionized species. One wonders then how the gas avoids full
photoionization and we see any spectral lines at all. Generally, two
mechanisms have been proposed to resolve the so-called overionizaton  
problem: (i) the AGN outflows have filling factors less than one
and consist of dense clouds and (ii)  the filling factors
equal one but the outflows are shielded from the powerful radiation 
by some material located between the central engine and the outflow
(e.g., Krolik 1999, and below).

Many theoretical models have been proposed to explain outflows
in AGNs.  Comprehensive reviews of recent theoretical work on outflows 
in AGNs can be found in Arav, Shlosman \& Weymann (1997). 
Several forces have been suggested to accelerate outflows in AGN, 
for example, gas pressure (e.g., Weymann et al. 1982; 
Begelman, de~Kool \& Sikora~1991), 
magneto-centrifugal force due to an accretion disk 
(e.g., Blandford \& Payne 1982; Emmering, Blandford \& Shlosman 1992; 
K$\ddot{\rm o}$nigl  \& Kartje 1994; Bottorff et al. 1997), 
radiation pressure due to dust (e.g., Voit, Weymann \& Korista 1993; 
Scoville \& Norman 1995),
and radiation pressure due to lines 
(e.g., Drew \& Boksenberg 1984; Shlosman, Vitello \& Shaviv 1985; 
Arav, Li \& Begelman 1994; de~Kool \& Begelman 1995, hereafter dKB; 
Murray et al. 1995, hereafter MCGV). 
Most of these models avoid  overionization of the wind
by assuming high density clouds.

One plausible scenario for the AGN outflows is
that radiation pressure on spectral lines drives a wind from an
accretion disk around a black hole. The presence of the BALs themselves
strongly indicates that substantial momentum is transfered from 
a powerful radiation field  to the gas. 
A crucial clue to the origin of AGN outflows comes from the discovery
of line-locking in BAL QSO spectra (e.g., Foltz et al. 1987).
Weymann et al. (1991)  discovered that a composite spectrum
of their BAL QSO sample shows 
a double trough in the C~IV $\lambda$1549 BAL, separated in velocity space
by the N V $\lambda$ 1240-Ly$\alpha$ splitting $\sim 5, 900$~km~$\rm s^{-1}$
(see also Korista et al. 1993). In fact Arav \& Begelman 
(1994, see also  Arav et al. 1995 and Arav 1996) showed 
that an absorption hump in the  C~IV $\lambda$ BAL is  ``the ghost
of Ly$\alpha$'' due to modulation of the radiation force by the
strong emission in Ly$\alpha$.

Our understanding of how line-driving produces powerful high velocity
winds is based  on the studies of winds in hot stars that radiate
mostly in the UV 
(e.g., Castor, Abbott, Klein 1975, hereafter CAK; Abbott 1982).  
The essential concept underpinning these models is that the momentum
is extracted most efficiently from the radiation field via line opacity.
With the inclusion of lines, CAK showed that the effective radiation force
can increase by several orders of magnitude above that due to 
electron-scattering alone. Thus even a star that radiates at around 0.1\%
of its Eddington limit, $L_{Edd}$ can  have a strong wind.
Up to now, however, it has been difficult to apply line-driven stellar
wind models to QSOs because of the fundamental difference in geometry:
stellar winds are to a good approximation spherically symmetric,
whereas the wind in BAL QSOs likely arises from a flattened
disk and is   therefore axially symmetric.

Line-radiation driven disk wind models for BAL QSOs have been proposed by 
several  authors.  For example, MCGV studied  a wind arising from a small 
distance from the central engine ($\sim 10^{16}$~cm).
Their model relies on the local disk radiation to launch gas from the disk 
and on the central radiation to accelerate the gas in the radial direction 
to high velocities. A key ingredient to this model is that the central engine 
radiation is attenuated by 'hitchhiking' gas located between 
the central engine and wind. Rather than showing from first principles 
the origin of this hitchhiking gas, they simply give plausible arguments 
for its existence. MCGV  adopted assumptions that require the flow to 
be time-independent and restrict the flow geometry to a quasi 1-D radial flow.
Other models of radiation-driven outflows in AGNs propose
that the radiation force accelerates the wind that has been
launched and kept at a low ionization state by {\it different} mechanisms.
For example, the dKB model supposes that
the wind is seated at a large distance ($\sim 10^{18}$~cm) from the central
engine. Because the disk is cool at these radii, 
the gas is not lifted from the disk by radiation pressure but by 
some alternative mechanism. Once the gas is high enough above the disk,
the radiation force due to the central engine accelerates the gas to high
velocity. The gas is not over-ionized in this model due to a small
filling factor caused by strong magnetic fields which confine the dense
clouds.

Recently it has been possible to model line-driven disk winds 
using 2-dimensional axisymmetric numerical hydrodynamical simulations
(e.g., Pereyra, Kallman \& Blondin 1997; 
PSD~I; Proga 1999; Proga, Stone \& Drew 1999, hereafter PSD~II). 
All these models have been calculated for white dwarf accretion disks.
PSD~I found that line-driven disk winds are produced only when the effective 
luminosity of the disk (i.e., the luminosity of the disk times the maximum 
value of the force multiplier, $L_D M_{max}$) exceeds 
the Eddington limit, $L_{Edd}$. If the dominant contribution to 
the total radiation field comes from the disk, then  the outflow 
is intrinsically unsteady and characterized by large amplitude velocity and 
density fluctuations. On the other hand, if the total luminosity of 
the system is dominated by the central object, then the outflow is steady.  
In either case, PSD~I and PSD~II found that the structure of 
the wind consists of a dense, slow outflow, that is bounded on the polar 
side by a high-velocity, lower density stream.  
The flow geometry is controlled largely by the geometry of 
the radiation field -- a brighter disk/central 
object produces a more polar/equatorial wind.  Global properties such
as the total mass loss rate and terminal velocity depend more on the
system luminosity and are insensitive to geometry.  The mass loss rate
is a strong function of the effective Eddington luminosity and is
of the same order of
magnitude as that of a simple spherically-symmetric stellar wind 
(see also Proga 1999).
Matter is fed into the fast stream from within a few central object radii.
The terminal velocity of the  stream 
is similar to that of the terminal velocity of a corresponding
spherical stellar wind, i.e., $v_\infty \sim {\rm a~few}~v_{esc}$,
where $v_{esc}$, is the escape velocity from the photosphere.

In this paper, we use the methods developed by PSD to study disk winds
in AGNs. However to study AGNs winds, we need to relax some of 
the assumptions adopted by PSD. In particular, we can no longer
assume the outflowing gas is isothermal, has a fixed ionization state, or
is optically thin. Instead, we adopt a simplified treatment of photoionization,
and radiative cooling and heating that allow us to compute self-consistently
the ionization state, and therefore line force, in the wind.

Here we calculate a few disk wind models using our extensions to the
PSD~II method. We concentrate on assessing how  winds
can be driven from a disk in the presence of very strong ionizing radiation.
We describe our calculation in Section 2. We present our results in Section 3
and discuss them together with perceived limitations in Section 4. 
The paper ends, in section 5, with our conclusions.

\section{Method}

\subsection{Hydrodynamics}

To calculate the structure and evolution of a wind
from a disk, we solve the equations of hydrodynamics
\begin{equation}
   \frac{D\rho}{Dt} + \rho \nabla \cdot {\bf v} = 0,
\end{equation}
\begin{equation}
   \rho \frac{D{\bf v}}{Dt} = - \nabla P + \rho {\bf g}
 + \rho {\bf F}^{rad} 
\end{equation}
\begin{equation}
   \rho \frac{D}{Dt}\left(\frac{e}{ \rho}\right) = -p \nabla \cdot {\bf v} 
   + \rho \cal{L}  ,
\end{equation}
where $\rho$ is the mass density, $P$ is the gas pressure, 
${\bf v}$ is the velocity, $e$ is the internal energy density,
$\cal{L}$ is the net cooling rate,
${\bf g}$ is the gravitational acceleration of the central object,
${\bf F}^{rad}$ is the total radiation force per unit mass.
We adopt an adiabatic equation of state
$P~=~(\gamma-1)e$, and consider models with $\gamma=5/3$.

Our calculations are performed in spherical polar coordinates
$(r,\theta,\phi)$. We assume axial symmetry about the rotational axis
of the accretion disk ($\theta=0^o$). However the $\theta=~90^o$ axis
is not coincident with the disk midplane. We allow this axis
to be above the midplane, at the height, $z_o$, corresponding
to the disk pressure scale height. 
Figure~1 shows a schematic representation of our computation domain. 
The offset enters our calculations
in the computation of the gravity; for example the gravity has non-zero
$\theta$ component at $\theta=90^o$, 
and in the computation of the Keplerian velocity
along the $\theta~=~90^o$ axis. 

This offset is introduced so that the bottom of our computational grid
is located at the disk photosphere, where the wind is launched. Unlike
the thin disk studied by PSD, the radiation dominated disk in AGN may
have significant geometrical width. If the base of the computational
grid is placed at the equatorial plane (midplane), this would require
modeling the full internal structure of the disk, which is complex and 
turbulent (e.g., Balbus \& Hawley 1998). Instead, we locate the base of the
grid where the vertical radiation force due to electron scattering 
cancels out the vertical component of  gravity. 

A realistic description of the radiation field from the central engine
would require detailed knowledge of the geometry of the flow near the
central engine, which is beyond the scope of this investigation. 
Instead we model the central engine as a point source
of radiation located at the origin of our computational grid
(that is located a height $z_o$ above the disk midplane). This implies
the central engine has  finite width, perhaps associated with a hot corona.
We expect the high column density of the disk to attenuate radiation
close to the $\theta=90^o$ plane.

Our standard computational domain is defined to occupy the radial range
$r_i~=~100~r_\ast \leq r \leq \ r_o~=~ 1000~r_\ast$, where $r_\ast$ is
the inner radius of the disk,  and the angular range
$0^o \leq \theta \leq 90^o$. For comparison, we also calculate some models
with the radial range $50~r_\ast \leq r \leq \ 1000~r_\ast$ 
and with the radial range $200~r_\ast \leq r \leq \ 1000~r_\ast$.
The $r-\theta$ domain is discretized into zones.  
Our  numerical resolution consists of 100
zones in each of the $r$ and $\theta$ directions, with fixed zone size
ratios, $dr_{k+1}/dr_{k}=1.05$ and $d\theta_{l}/d\theta_{l+1} =1.087$.
Gridding in this manner ensures good spatial resolution close to 
the radiating surface of the disk and the inner boundary at $r_i$.

The boundary conditions are specified as follows. At
$\theta=0$, we apply an axis-of-symmetry boundary condition.  For
the outer radial boundary, we apply an outflow boundary 
condition.  For the inner radial boundary $r=r_{\ast}$
and for $\theta=90^o$, we apply reflecting boundary conditions for the density,
velocity and internal energy.
To represent steady conditions in the photosphere at the base of the wind, 
during the evolution of each model we apply the constraints that in
the first zone above the $\theta=90^o$ plane the radial velocity $v_r=0$,
the rotational velocity $v_\phi$ remains Keplerian, and the density is
fixed at $\rho = \rho_0$ at all times.  
The initial conditions are as in PSD~I expect for  the initial temperature
profile which is described in detail in Section 2.3.

To solve eqs. 1-3 we use an extended version of the ZEUS-2D code 
(e.g., Stone \& Norman 1992). The equation for the internal energy, eq. 3,
is solved  using the operator splitting method and the backward Euler scheme. 

\subsection{The radiation field and force}

The geometry and assumptions needed to compute the radiation
field from the disk and central object are as in PSD~II (see also PSD~I).
The disk is flat, Keplerian, geometrically-thin and
optically-thick.  
The disk photosphere is  coincident with the $\theta=90^o$ axis.
We specify the radiation field of the disk  by assuming that the temperature 
follows the radial profile of the optically thick accretion disk 
(Shakura \& Sunyaev 1973), and therefore depends on
the mass accretion rate in the disk, $\dot{M}_a$, the mass of the black hole, 
$M_{BH}$  and  the inner edge of the disk, $r_\ast=3 r_S$, where 
$r_S=2GM_{BH}/c^2$ is the Schwarzschild radius of a black hole. 
In particular, the disk luminosity, $L_D=2 \eta G M_{BH} \MDOT_a/r_S$,
where $\eta$ is the rest mass conversion efficiency. 

The geometry of the central engine 
in AGNs is not well known. For simplicity, we consider the central engine 
as the most inner part of the accretion disk plus an extended corona.
We refer to the corona as the central object. 
The radius of the central object is comparable with the inner radius of the 
disk and we assume  that they formally are equal.

We express the central object 
luminosity $L_\ast$ in units of the disk luminosity $L_\ast=x L_D$.
In contrast to PSD~I and PSD~II, we allow for the situation
when only some fraction of the central object luminosity 
takes part in  driving a wind. We identify this fraction 
as the luminosity in the UV band, $f_{\rm UV}L_\ast$. 
We refer to the fraction of the luminosity that is responsible 
for  ionizing  the wind to a very high state as the luminosity in 
the X-ray band, $f_{\rm X}L_\ast$. For simplicity, we assume here
that this fraction of the luminosity does not contribute to line driving of 
the wind. 
We call the luminosity in the remaining bands, mainly optical and infrared, 
as $f_{\rm O,IR} L_\ast$. We assume that $f_{\rm O, IR}L_\ast$
is the part of the luminosity that does not change the dynamics of the wind. 
We stress that the fraction $f_i$ is a numerical factor that
we introduce here to parameterize the luminosity in each of those three
domains. We set $f_{\rm OPT,IR}$ to zero in the remaining part of the paper
as the central object is very hot.
We take into account the irradiation of the disk by the central object, 
assuming that the disk re-emits all absorbed energy locally 
and isotropically. 
However we note that the contribution from irradiation is negligible 
for $x\sim 1$ and large radii (see eq. 6 below).

We approximate the radiative acceleration due to lines
(line force, for short) using a modified CAK method.  The line force
at a point defined by the position vector $\bf r$ is
\begin{equation}
{\bf F}^{rad,l}~({\bf{r}})=~\oint_{\Omega} M(t) 
\left(\hat{n} \frac{\sigma_e I({\bf r},\hat{n}) d\Omega}{c} \right)
\end{equation}
where $I$ is the frequency-integrated continuum intensity in the direction
defined by the unit vector $\hat{n}$, and $\Omega$ is the solid angle
subtended by the disk and central object at the point. 
The term in brackets is the electron-scattering radiation force,
$\sigma_e$ is  the mass-scattering coefficient for free electrons,
and $M(t)$ is the force multiplier -- the numerical factor which
parameterizes by how much spectral lines increase the scattering
coefficient. In the Sobolev approximation, $M(t)$ is a function
of the optical depth parameter
\begin{equation}
t~=~\frac{\sigma_e \rho v_{th}}
{ \left| dv_l/dl \right|},
\end{equation}
where $v_{th}$ is the thermal velocity, 
and $\frac{dv_l}{dl}$ is the velocity gradient along the line of sight, 
$\hat{n}$.

We evaluate the radiation force in four steps. First, we calculate the
intensity, the velocity gradient in the $\hat{n}$ direction and then the 
optical depth parameter $t$. Second, we calculate the parameters of the
force multiplier using a current value of the photoionization
parameter, $\xi$ adopting results of Stevens \& Kallman (1990). 
Then we calculate the radiation force exerted
by radiation along $\hat{n}$. Third, we integrate the radiation
force over the solid angle subtended by the radiant surface. 
Finally, we correct the radiation force in the radial direction
for the optical depth effects.

Our numerical algorithm
for evaluating the line force for given parameters of the 
force multiplier, the third step, is described in PSD~II.  
For a rotating flow, there may be an azimuthal component to the line force 
even in axisymmetry. However we set this component of the
line force to zero because it is always less than  other components
(e.g., PSD~II). 
See PSD~I and PSD~II for further details.
Below we describe our calculations of the force multiplier for 
various conditions in the wind and our treatment of the  optical
depth effects on the radiation force.

In the disk plane at $r=r_{D}$, $I({\bf r}, \hat{n})$ is the local
isotropic disk intensity:
\begin{eqnarray}
\rule{0in}{3.0ex}
I_{D}(r_D) & =~\frac{3 G M_{BH} \MDOT_a}{8 \pi^2 r_\ast^3} 
\left\{ \rule{0in}{3.0ex}  \frac{r_\ast^3}{r_D^3}\left(1 - 
\left( \frac{r_\ast}{r_D}\right)^{1/2}\right) \nonumber \right.~~~~~~~~~~~~~\\
 & \left. +\frac{x}{3\pi}\left(\arcsin \frac{r_\ast}{r_D} - 
\frac{r_\ast}{r_D} \left(1 - 
\left(\frac{r_\ast}{r_D}\right)^2\right)^{1/2}\right) \right\}.
\end{eqnarray}
We consider only the hot part of the disk, where the local disk temperature, 
$T_D \simgreat$ a few  $10^3$~K. We then assume that all disk photons 
can contribute to the line force. 
The second term in the curly brackets corresponds to the contribution from the
irradiation of a disk by a central object. For large radii and
$x \sim 1$, the contribution from the irradiation is negligible compared
to the intrinsic disk intensity (the first term in the curly brackets).

For radiation from the central object, the intensity may be written as:
\begin{equation} 
I_\ast~=~\frac{L_\ast}{4\pi^2 r_\ast^2}.
\end{equation}
We account for the fact that the central radiation consists of the two
distinct spectral components by using the parameter 
$f_{i}$,  where $i={\rm X}$ or ${\rm UV}$. 
We assume that the optical depth effects in the $\theta$ direction 
are negligible for the central radiation in the two bands. Note 
that the $\theta$  component of the radiation force due to 
the central object is negligible in our computational domain because 
the domain is far from the object ($r_\ast \ll r_i$).
In the radial direction however, the column density can be high
and the optical depth effects can be significant.
We estimate the optical depth, $\tau_{\rm i}$ between the central  
source and a point in a wind from:
\begin{equation}
 \tau_{\rm i}= \int_0^{r} \kappa_{\rm i} \rho dr, 
\end{equation} 
where $\kappa_{\rm i}$ is the absorption coefficient
representative for the $i$ band,
and $r$ is the distance from the central source.
We can treat the central object as a point source because
our calculations are for $r_\ast/r \ll 1$. 
Then we calculate the radial component of the central radiation force 
due to lines  from
\begin{equation}
({F}^{rad,l}_\ast)_{r}~({\bf{r}})=~f_{\rm UV} \exp(-\tau_{\rm UV})
\oint_{\Omega_\ast} M(t) 
\left(n_r \frac{\sigma_e I_\ast({\bf r},\hat{n}) d\Omega}{c} \right),
\end{equation}
where $\Omega_\ast$ is the central object solid angle.
We also take into account the optical depth effects on the radial 
component of the electron-scattering  force:
\begin{equation}
({F}^{rad,e}_\ast)_{r}~({\bf{r}})=~\{ f_{UV} \exp(-\tau_{\rm UV})+
f_{\rm X} \exp(-\tau_{X}) \}
\oint_{\Omega_\ast}  
\left(n_r \frac{\sigma_e I_\ast({\bf r},\hat{n}) d\Omega}{c} \right).
\end{equation}
The attenuation of the X-ray radiation is calculated using  
$\kappa_{\rm X}=40~{\rm g^{-1} cm ^2}$ for $\xi \leq 10^5$ and  
$\kappa_{\rm X}=0.4~{\rm g^{-1} cm ^2}$ for $\xi > 10^5$, while
the attenuation of the UV radiation is calculated using
$\kappa_{\rm UV}= 0.4~{\rm g^{-1}~cm^2}$ for all $\xi$.

A disk wind may also be optically thick to the UV radiation emitted by 
the disk. 
In particular, the flux along the radial direction near the 
disk can be significantly reduced due to  high column density.  
We approximate this effect by multiplying the radial
component of the disk radiation force by the attenuation factor 
$\exp(-\tau_{\rm UV})$.


To calculate the force multiplier, we adopt the CAK analytical expression  
modified by Owocki, Castor \& Rybicki (1988, see also PSD~I)
\begin{equation}
M(t)~=~k t^{-\alpha}~ 
\left[ \frac{(1+\tau_{max})^{(1-\alpha)}-1} {\tau_{max}^{(1-\alpha)}} \right]
\end{equation}
where $k$ is proportional to the total number of lines, $\alpha$ is the 
ratio of optically-thick to optically-thin lines, 
$\tau_{max}=t\eta_{max}$ and $\eta_{max}$ is a parameter determining
the maximum value, $M_{max}$ achieved for the force multiplier.  
Equation 11 shows the following limiting behavior:
\begin{eqnarray}
\lim_{\tau_{max} \rightarrow \infty}~M(t) & = & k t^{-\alpha} \\
\lim_{\tau_{max} \rightarrow 0}~M(t) & = & M_{max},
\end{eqnarray} 
where $M_{max} = k (1-\alpha)\eta_{max}^\alpha$. The maximum value of the 
force multiplier is a function of physical parameters of the wind and
radiation field. Several studies have showed that $M_{max}$ is roughly
a few thousand for gas ionized by a weak or moderate radiation field
(e.g., CAK; Abbott 1982; Stevens \& Kallman 1990; Gayley 1995).
As the radiation field becomes stronger and the gas becomes more ionized
the force multiplier decreases asymptotically to zero.

The line force, in particular the parameters of the force multiplier
depend on the ionization of the wind and the spectral energy distribution
(SED) of the radiation field.
Self-consistent calculations of the line-force for given wind conditions  
and the radiation 
require detailed calculations of the wind photoionization structure (e.g.,
CAK; Abbott 1982; Puls et al. 2000).
The  total line force includes contributions from  $>10^5$ lines 
from many species. Such calculations in connection with the 2-D, 
time-dependent hydrodynamic calculations are not feasible. 
We therefore start by adopting the analytical formulae
for the force multiplier due to Stevens \& Kallman (1991). 
They studied the effects of X-ray ionization on the the  radiative force
experienced by the stellar wind in a massive X-ray binary (MXRB).
Wind conditions in MXRBs differ somewhat from those in QSOs.
In particular, the X-ray radiation in MXRBs can be well represented by   
a 10~KeV bremsstrahlung spectrum while the spectrum in QSOs is better
fitted by  power laws with different spectral indices at different
spectral bands (e.g., Zheng et al. 1997; Laor et al. 1997). 
Nevertheless we adopt the Stevens \& Kallman results as a first order 
approximation because the Compton temperature and the ionization
structure are similar (Kallman \& McCray 1982).

Using the photoionization code XSTAR, Stevens \& Kallman 
found that the line force
due to the radiation from the primary can be parameterized
in terms of the CAK force multiplier. Their results show that the line
force decreases sharply with increasing photoionization parameter:
\begin{equation}
\xi = \frac{4 \pi {\cal F}_{\rm X}}{n},
\end{equation}
where ${\cal F}_{\rm X}$ is the local  X-ray flux, $n$
is the number density of the gas ($={\rho}/({m_p \mu })$ , where  
$m_p$ is the proton mass, and $\mu$ is the mean molecular weight).
Stevens \& Kallman also found some simple analytical fits to their results 
that  allowed them to express
$k$ and $\eta_{max}$ as functions of $\xi$:
\begin{equation}
k= 0.03 + 0.385 \exp(-1.4\xi^{0.6}),
\end{equation}
and 
\begin{equation}
\log_{10} \eta_{max} = \left\{ \begin{array}{ll} 
6.9~\exp(0.16~\xi^{0.4})  
& {\rm for}~~\,~~ 
\log_{10} \xi~\leq~0.5 \\
 & \\
9.1 \exp(-7.96~\times10^{-3}\xi)
& {\rm for} ~~\,~~
\log_{10}~\xi~>~0.5  \\
\end{array}
\right.
\end{equation}
The  parameter $\alpha=0.6$ and does not change with $\xi$. 

These expressions for the parameters of the force multiplier 
predict that $M_{max}$ increases gradually from $\sim 2000$
to $5000$ as $\xi$ increases from 0 to $\sim 3$ and then drops to $\sim 0.1$
at $\xi=1000$. The line force becomes  negligible
for $\xi \simgreat 100$ because then $M_{max} \sim 1$. 

To calculate the photoionization parameter, we first
need to know the local X-ray flux ${\cal F}_{\rm X}$ (see eq.~14).
We estimate  ${\cal F}_{\rm X}$ assuming that the source of all X-rays 
is a point-like central object with the luminosity in the X-ray band, 
$L_{\rm X}=f_{\rm X}L_\ast$. 
Making the above simplifications, we can express the local X-ray flux as
\begin{equation}
{\cal F}_{\rm X}=\exp(-\tau_{\rm X})\frac{L_{\rm X}}{4 \pi r^2}.
\end{equation} 

\subsection{Radiation heating and cooling}

We calculate the gas temperature  assuming that the gas 
is optically thin to its own cooling radiation and 
that the abundances are cosmic 
(Withbroe 1971).
The net cooling rate depends 
on the  density, $\rho$, the temperature, $T$, the ionization parameter 
$\xi$, and the characteristic temperature of the X-ray radiation $T_{\rm X}$.
It is then possible to fit analytical formulae to
the heating and cooling rate obtained from  detailed  photoionization 
calculations for various $T_{X}$, $T$, and $\xi$. 
For example, Blondin (1994) found that for a 10 keV bremsstrahlung spectrum, 
his fits typical agree with a detailed calculation (Blondin et al. 1990) 
to within 25\%. Using Blondin's result we can express the net cooling rate
$\cal {L}$ in equation 3  by 
\begin{equation}
\rho {\cal L}= n^2 
(G_{Compton}+G_{\rm X}-L_{b,l})~erg~\rm cm^{3}~s^{-1},
\end{equation}
where $G_{Compton}$ is the rate of Compton heating/cooling,
\begin{equation}
G_{Compton}~=8.9\times10^{-36}~\xi~(T_{\rm X}-T)~\rm~erg~\rm cm^{-3}~s^{-1},
\end{equation}
$G_{\rm X}$ is the net rate of X-ray photoionization heating--recombination
cooling 
\begin{equation}
G_{\rm X}~=~1.5~\times10^{-21}~\xi^{1/4}~T^{-1/2}(1-T/T_{\rm X})
~\rm~erg~\rm cm^{-3}~s^{-1}
\end{equation}
and $L_{b,l}$ is the rate of bremsstrahlung and line cooling
\begin{equation}
L_{b,l}=3.3~\times~10^{-27}~T^{1/2}+[1.7~\times~10^{-18} 
\exp(-1.3~\times~10^5/T)\xi^{-1}T^{-1/2}+10^{-24}] \delta
~\rm~erg~\rm cm^{-3}~s^{-1}.
\end{equation}
The parameter $\delta$ is introduced to control line cooling, 
$\delta~=~1$ represents optically thin cooling and $\delta < 1$ 
represents the case when lines become optically thick and 
cooling is reduced.

As we mentioned in Section 2.1, we use the same initial and 
boundary conditions as in PSD~I.
However we here also calculate the evolution of the internal energy.
We  specify the boundary  conditions and the initial conditions
for $e$ as follows. For the temperature at the 
base of the wind, in the first grid zone above the $\theta=90^o$ 
plane, we adopt the solution for 
the steady state disk.
Adopting the assumption that  the local disk intensity is 
a black body and using eq. 6, the disk temperature is:
\begin{equation}
T_D(r_D)= (\pi I_D(r_D)/\sigma)^{1/4}.
\end{equation} 
Then we calculate the internal energy from 
\begin{equation}
e_o(r_D)=\frac{\rho_o k T(r_D)}{\mu m_p (\gamma -1)},
\end{equation}
where $\rho_o$ is the density in the first grid zone above the 
$\theta=90^o$ plane. We consider models with $\mu=1$.
We initialize the internal energy at other locations by assuming that
the gas temperature at a given radius is the same as at the disk
temperature (i.e., $T(r, \theta)= T_D(r \sin \theta)$ for all $\theta$) and 
using initial value of the density (see PSD~I).
Consistent with the boundary conditions  for 
other gas quantities (see Sect.~2.1) we apply the constraint that 
in the first zone above the $\theta=90^o$  plane the internal energy 
$e=e_o$ at all times. We find only the initial, transient evolution
of the wind is affected by this choice of initial temperature conditions.

Our treatment of radiative heating and cooling is most likely valid
in a low density and high photoionization parameter regime.
For a high density and low photoionization parameter regime
our treatment will likely underestimate the temperature because 
we oversimplify line cooling by not properly taking into
account optical depth effects. In dense regions, our scheme therefore
may yield a very low gas temperature (e.g., $T \ll 1000 $~K). In such
cases we set the lower limit for the temperature assuming that the gas
is in  local thermodynamical equilibrium, i.e., the gas temperature
equals the local disk radiation temperature. Specifically, 
if at the ($r, \theta$)  wind point
the temperature from eq. 3 is less than $T_D (r sin \theta)$
then we replace it with $T_D (r sin \theta)$.

\section{Results}

We specify our models by several parameters. 
In all our calculations we assume the mass of the non-rotating black hole,  
$M_{BH}=10^8~\rm \MSUN$ and the rest mass conversion efficiency, $\eta=0.06$
typical for AGN.
To determine the radiation field from the disk, we assume the mass accretion
rate $\MDOT_a=1.8$~M$_{\odot}$~yr$^{-1}$. 
These system parameters yield the disk Eddington number, $\Gamma_D=0.5$,
the disk inner radius, $r_\ast=8.8\times10^{13}$~cm   
and the orbital period at the inner edge of the disk, 
$\tau= \sqrt{\frac{r_\ast^3}{G M_{BH}}}= 7.22\times 10^{3}$ sec.
We set the offset of the computational domain, $z_o~= H_D \approx 3 \Gamma_D$
that corresponds to the half width of the radiation pressure dominated disk
at large radii (e.g., Shakura \& Sunyaev 1973).
The radiation field from 
the central engine is specified by the additional parameters: $x=1$, 
$f_{\rm UV}=0.5$, $f_{\rm X}=0.5$.
As we discussed above the SED of the ionizing radiation is not
well known, our choice of values for $f_{\rm UV}$ and $f_{\rm X}$
is guided by the results from  Zheng et al. (1997) and Laor et al. (1997).
To calculate the line force we 
adopt the force multiplier parameter $\alpha=0.6$. We calculate the 
remaining parameters of the multiplier, i.e., $k$ and $\eta_{max}$ 
as functions of  the photoionization parameter, $\xi$ using  eqs 15 and 16. 
The force multiplier depends only formally on the thermal speed, 
$v_{th}$ which we set to 20 ${\rm km~s^{-1}}$, i.e., the thermal speed of 
a hydrogen atom at the temperature of 25000~K (Stevens \& Kallman 1990).  
To calculate the gas temperature, we assume the temperature of 
the X-ray radiation, $T_{\rm X}=10$~KeV and the line cooling parameter 
$\delta=1$.

We adopted a generalized CAK method to calculate a disk wind driven by 
the line force (PSD~II). The CAK method has been developed for OB stars with
photospheres dominated by gas pressure (the inner atmosphere is 
in hydrostatic equilibrium), and they radiate most of their energy
in the UV band. We chose the radial extent of the computational 
domain (i.e., $100~r_\ast \leq r \leq 1000~r_\ast$) keeping in mind  
that the disk atmosphere should satisfy these two conditions, however
some compromises had to be done. 
For example, the disk temperature at  $100~r_\ast=8.8\times10^{15}$~cm 
is 8200~K 
while at $1000~r_\ast=8.8\times10^{16}~\rm cm$ is 1500~K using our disk 
parameters. We realize that according to the $\alpha$-disk model 
the disk emits mostly in the UV band down to the radius of 
$\sim 10~r_\ast$ where its effective temperature is $50000$~K.
However the disk at this small radius is likely radiation dominated
(e.g., Shakura \& Sunyaev 1973; Svensson \& Zdziarski 1994).  
On the other hand, the disk atmosphere at radius $1000~r\ast$
and the effective temperature of $1500~K$, radiates most of its energy 
in the infrared band. Nevertheless
we chose a large value for the outer edge of the computational domain to
make sure that the domain includes most of the acceleration zone of the disk
wind launched from the inner disk and the wind velocity at the outer
edge well represents the wind terminal velocity.

Figure~2 shows the instantaneous density, temperature and 
photoionization parameter distributions and the poloidal velocity field
of the model.
After $12.6$~years ($\sim~5.5\times 10^4~\tau$), 
the disk material fills the grid for $\theta \simgreat 70^o$
and remains in that region for the rest of the run, i.e.,
over next $47.7$~years. Figure~2
shows results after $14.6$~years.  
Although the flow is time-dependent 
the gross properties of the flow (e.g., the mass loss rate and the radial
velocity at the outer boundary), settle down to steady time-averages
over timescales on the order of $3$~years.
Our calculation follows
(i) a  hot, low density flow in the polar region
(ii) a dense, warm and fast equatorial  outflow from the disk, (iii)
a transitional zone in which the disk outflow is hot and struggles to
escape the system.

In the polar region, the density is very small and close to the lower
limit that we set on the grid, i.e., $\rho_{min}=10^{-20}~{\rm g cm^{-3}}$.
The line force is negligible because the matter is highly ionized 
as indicated by a very large photoionization parameter ($\sim 10^7$). 
The gas temperature is close to the temperature of the X-ray radiation,
again indicative of highly ionized gas. 
The matter in the polar region is pulled by the gravity 
from the outer boundary and it is an artifact of the boundary conditions.
Overall this region of the very low density is not relevant 
to our analysis as it has no effect on the much denser disk flow.

The equatorial region is distinctly different. In the inner part
of the disk (i.e., for $r \sim r_i$), 
the density at the wind base is high, $\sim10^{-13}~{\rm g~cm^{-3}}$.
Thus the photoionization parameter is low  despite the strong central 
radiation. However as the flow from the inner part of the disk is accelerated 
by the line force its density decreases and the gas temperature and 
the  photoionization parameter increase. Subsequently the gas becomes fully 
ionized and loses all of driving lines before it  reaches the escape velocity
and therefore falls on the central object/inner disk.
Although this gas does not produce a wind, its primary effect is to shield
the gas at larger radii. Thus the wind consists of gas accelerated by 
the line force at larger radii, in fact driving of the disk wind extends 
over all radii at which the intrinsic disk radiation is large enough to 
launch gas. 

The poloidal velocity  (Fig. 2c) shows that the gas streamlines are 
perpendicular to the disk over some height that increases with radius.
The streamlines then bend away from the central object and converge. 
The region where the flow is moving almost radially outward is associated 
with a high-velocity, high density stream. This fast stream contributes 
$\sim$ 100\% to the total mass loss rate, 
$\MDOT_W= 0.5~\rm \MSUNYR$. We note that the mass loss rate can increase
by a factor of a few when a knot is crossing the outer boundary.

The fast stream is variable. 
In the upper envelope of the disk wind there is large velocity shear between
the higher density fast stream moving outward and the lower density
fast gas moving inward.
Our simulations show that this shear gives rise to Kelvin-Helmholtz 
instabilities. The instabilities generate knots that propagate
along the fast stream at  $\sim 10000$~km~$\rm s^{-1}$. 
Figure~2a shows an example
of such a knot at $r\sim 750~r_\ast$ and $z \sim 200~r_\ast$,
in the figure coordinates.
The density contrast between the knot and the fast stream is $\sim 2$
orders of magnitude.
The knots generator is episodic and is inherent to the fast stream.
We observed the generation of 19 knots over $60.4$~years. 
In other words, a knot is produced every $\sim 3$ years. 

Figure~3 presents a sequence of density maps showing time evolution
of the outflow from Figure~2 after 13.3, 14.6 and 16.47 years, 
left, middle and righ panel respectively. Note that the middle panel
from Figure~3 is the same as the top left panel from Figure~2.
Figure~3 well illustrates variability in the outflow, in particular,
generation of a knot in the fast stream at 
$r\sim 450~r_\ast$ and $z \sim 100~r_\ast$ (left panel), 
the well-formed knot at $r\sim 750~r_\ast$ and $z \sim 200~r_\ast$ 
(middle panel), and the time when the knot left the gird  at
$r\sim 950~r_\ast$ and $z \sim 300~r_\ast$,
and the fast stream is fairly smooth in between episodes with knots 
(right panel).
The time dependence in the region close to the disk,
bounded by the fast stream resembles outward propagation of a wave
at the begin of evolution but becomes more complex with time as elements 
of the flow move both upwards and downwards.

Figure~4 presents the run of the density, radial velocity, photoionization
parameter and column density as a function of the polar angle, $\theta$
at the outer boundary, $r_o=8.8\times10^{16}$~cm from Figure~2. 
The column density is given by:
\begin{equation}
N_H(\theta)= \int_{r_i}^{r_o} \frac{\rho(r, \theta)}{\mu m_p } dr.
\end{equation}
The gas density is a very strong function
of angle for $\theta$ between ~$90^o$ and $65^o$.  
The density 
drops by $~ 8$ orders of magnitude between $\theta = 90^o$ and 
$\theta \sim 89^o$,  as might be expected
of a density profile determined by hydrostatic equilibrium.
For $70^o \simless \theta \simless 89^o$, the wind domain, $\rho$ varies 
between $10^{-19}$ and $10^{-17}~\rm g~cm^{-3}$.  For $\theta \simless 70^o$, 
density  decreases gradually to so low a value that it becomes necessary 
to replace it by the numerical lower limit $\rho_{min}$.
The radial velocity at 
$1000~r_\ast~=~8.8\times10^{16}$~cm 
has a broad peak for $72^o \simless \theta \simless 82^o$ with the maximum
of ~15000~km~$\rm s^{-1}$ at $\theta\sim 75$. On both side of the peak
the velocity is close to zero.

The photoionization parameter is very high, $\sim 10^7$, for 
$\theta \simless 70^o$ because of the very low density. However
it drops by ~15 orders of magnitude between 
$70^o \simless \theta \simless 75^o$ and 
stays at a very low level for $\theta > 75^o$.
The column density changes less dramatically with angle.
Between the pole and $\theta \simless 67^o$, $\rm N_H$ is less than 
$10^{22}~\rm cm^{-2}$.
For $\theta > 67^o$, $\rm N_H$ increases gradually with $\theta$,
it reaches value of $10^{24}~\rm cm^{-2}$ at $\theta \sim 80^o$. For 
$80^o \simless \theta \simless 89^o$, $\rm N_H$ varies
between $10^{24}~\rm cm^{-2}$ and $10^{26}~\rm cm^{-2}$.

Our model shows that (i) the intrinsic disk radiation can launch a wind
and (ii) the central object radiation can accelerate the disk wind 
to very high velocities. We have checked how these main results are sensitive
to the model parameters. 
For a fixed disk atmosphere and central radiation source, the most important
parameter of our model is the optical depth for the X-rays from the
central object. Therefore we have run 
several tests with various X-ray opacities for $\xi < 10^5$: 
$\kappa_X=0.4~{\rm g^{-1}~cm^2}$ and $\kappa_X=4~{\rm g^{-1}~cm^2}$. 
The former case is most conservative
because it corresponds to the case when
the X-ray optical depth is reduced to the Thomson optical depth.
We found that in both test runs  the X-ray radiation is significantly
attenuated so the line force can launch a disk outflow.
However  the flow velocity never exceeds the escape velocity 
because the X-rays fully ionize the gas  close to the disk and 
produces a hot corona with complex velocity field but which does not
escape the system. 
We conclude then that the disk atmosphere can 'shield' itself at least
to the extent that the local disk radiation can launch gas off the disk
photosphere.

The two test runs also illustrate how robust is our second result:
the central radiation accelerates the wind to very high velocities.
For this to happen the column density, $N_{\rm H}$
must be large enough to reduce the X-ray radiation
but too not large to reduce the UV flux in the radial direction. 
In other words, this requires $\tau_{\rm ~X}> 1$
and at the same time, $\tau_{UV} \ll 1$.

The test run with $\kappa_{\rm X}=0.4~{\rm g^{-1}~cm^2}$ corresponds
to the situation where $\tau_{\rm X}=\tau_{\rm UV}$. Thus the disk gas
that is sufficiently shielded from the X-rays
is  also shielded from the UV photons from the central object.
Such gas can only be lifted from the disk surface by the disk UV radiation
but fails to gain momentum in the radial direction.
 
The fact that the disk wind can be launched without {\it external} 
shielding material implies that our solution can depend on the location of 
the inner computational radius, $r_i$.
We have run several tests   
with $r_i=50r_\ast$ and  $r_i=200r_\ast$ to check this (the remaining
model parameters were as in the models shown in Figure~1 and~2).
We found that the location of the inner edge, $r_i$ affects the properties of 
the wind but not the fact the wind is produced. 
For example,
the wind opening angle, $\omega$, is $25^o$, $15^o$ and $8^o$
for $r_i=50, r_\ast, 100 r_\ast$ and $200 r_\ast$, respectively.
The maximum radial velocity at the outer boundary  decreases
from 20000~$\rm km~s^{-1}$ through, 15000~$\rm km~s^{-1}$ to
5000~$\rm km~s^{-1}$ for above sequence of decreasing $r_i$.
The mass loss rate also decreases with $r_i$ from 0.6~$\MSUNYR$
through 0.5~$\MSUNYR$ to 0.2~$\MSUNYR$.

\section{Discussion}

Our calculations show that the UV emitting accretion disk can launch
a wind that will shield itself from the strong ionizing radiation
emitted from the central object. The self-shielding is quite
robust because the external radiation does not penetrate the disk
down to the region where the line force becomes important and pushes
matter away from the photosphere. We illustrated the robustness of 
the self-shielding by changing  relevant model parameters.
We also note that a similar effect was observed by Stevens \& Kallman (1990) 
and Stevens (1991). They studied the effects of X-ray ionization on 
the radiative force experienced by the stellar wind in a MXRB.
They calculated the detailed photoionization structure of the line-driven 
wind for various X-ray luminosities without optical depth effects
(Stevens \& Kallman 1990) and with optical depth effects (Stevens 1991).
Additionally, they considered irradiation at various directions including
the normal to the mass losing photosphere. In none of the models
the irradiation penetrated below the so-called wind critical point
where, in the CAK type models, the wind mass loss rate is determined.  
They found, as we did, that the irradiation can severely decrease
the wind velocity in the supersonic portion of the flow.

We model here a line-driven wind from a disk that is flat, Keplerian, 
geometrically-thin and optically-thick. For the temperature at the base of 
the wind, in the first grid zone above the $\theta=90^o$ plane, we adopt 
the solution for the steady state $\alpha$ disk. We calculate the local disk 
intensity  as if it emits  as a black body. In all our calculations, 
we treat the regions for $\theta \sim 90^o$ as if they are gas pressure 
dominated. We find  that our results depend on  the inner radius of 
the computational domain. This dependence is likely  an artifact of one or 
more of our assumptions. For example, for very high luminosities the inner 
part of the disk is dominated by the radiation pressure. For the parameters 
adopted here, namely $\Gamma=0.5$, the inner radius of a gas pressure 
dominated disk is $\sim 10^{16}$~cm (e.g., Svensson \& Zdziarski 1994).  
The structure of the radiation dominated disks is not well known.
We  expect, however, that incorporating the radiation dominated part of 
the  disk into our calculation will change the condition of the base of 
the wind, for example, the gas density and  the UV flux might be reduced. 
These changes might affect the solution for the wind, in particular, the wind 
mass flux. We plan to explore these issues in a future paper.

To calculate the disk surface temperature and intensity, we took into account 
the irradiation of the disk by the central object.
However the contribution from the irradiation is negligible
compared to the disk intrinsic temperature because we considered 
a flat disk at large radii. 
Additionally, the high column density of the disk wind,
that we found in our models, implies that
the irradiating flux will be significantly attenuated  by the wind
before it reaches  the disk surface.
We anticipate that the last effect will significantly reduce 
disk irradiation  regardless of the shape of the disk surface -- 
flat or flaring. Detailed NLTE photoionization
calculations  are required to determine what fraction
of the central radiation will reach the surface of 
a mass losing disk. A complete treatment of disk irradiation is difficult
also because it is not obvious a priori whether the radiation incident on 
the disk  is completely  thermalized and re-radiated isotropically or 
it is scattered off the disk atmosphere or both.  

Our calculations show that  a line-driven disk wind model offers
a  promise of explaining outflows in AGNs. However they also 
illustrate some problems with this model. For example, as dKB pointed out in 
discussing MCGV's model, a very small radius at which the disk wind is 
launched also implies a very small size to BELR because BELR lies inside or 
is cospatial with BALR. The small scales in turn imply short crossing time of 
the BALR, of order of a month or so, and it is difficult to understand 
that the highly complex kinematical structures that BALs often exhibit do not 
appear to vary on timescales of 10 years (Barlow 1994). 

Our axisymmetric 2D models show that the disk wind is unsteady
and generates dense knots every 3~years. These knots will 
correspond to rings in 3D.
However in fully 3D calculations the rings may break down
to spirals or clouds or both. The breaking down of the rings will
change the density contrast between them and the rest of the wind.
Thus it is not clear if we would be able to see any spectral signature
of density fluctuations in the wind. Detail calculations of line profiles
are required to check this point.
Additionally, detailed photoionization calculations 
are required to check if the full range of ions observed to show 
BAL profiles can be explained:
BALs from ions with ionization potentials as low as of O~III or lower and 
as high as of O~VI.

To produce a fast wind the ratio between $\tau_{\rm X}$ to $\tau_{\rm UV}$ 
is very important. The low ratio gives a slow disk wind whereas the high
ratio gives a fast wind. This result is consistent with 
the observed anti-correlation for QSOs between 
the relative strength of the soft X-ray flux and the CIV
absorption equivalent width (e.g., Brandt, Laor \& Wills 2000).
However there is an
upper limit for the X-ray attenuation, or the column density between
the X-ray source and wind, namely for $N_H\simgreat 10^{24}~{\rm cm^{-2}}$
($\tau_{\rm X}>> 1$) 
the gas  is well shielded from the X-rays but at the same time
it is also shielded from any other radiation from the central engine
including the  UV radiation. In this case, the line force
from the central engine could be so much reduced that it may not
accelerate the gas  to high radial velocities and no strong wind will
be produced.


Some of the emission-line properties of QSOs could be explained as 
a correlation between luminosity and the slope of the ionizing spectrum,
i.e., lower luminosity objects have harder spectra (e.g.,
Boroson \& Green 1992).
Using our parameterization of the AGN radiation, this QSO's property
can be represented by increasing $f_{\rm X}$ in expense of $f_{\rm UV}$
when we reduce the luminosity. Such a choice of model parameters
will increase the strength of X-ray ionizing radiation in comparison to
the UV driving radiation. Consequently a disk wind should be weaker and
slower as some of our test runs indicate. 

Boroson \& Green (1992) argue that the dominant source of variation
in the observed properties of low redshift QSOs is not driven by external
orientation but rather by the fraction of the Eddington luminosity at
which the object is emitting (our parameter $\Gamma_D$).
We plan to examine the parameter space of our models to define the major
trends in disk wind behavior that will help us to explain
the observational trends found in various AGNs.

\section{Conclusions}

We have studied radiation driven winds from luminous accretion 
disks using numerical methods to solve the two-dimensional, time-dependent
equations of hydrodynamics. In so doing we have accounted for the radiation
force due to spectral lines using a generalized multidimensional formulation
of the Sobolev approximation. Additionally we have taken into account
the effects of the  strong central radiation on the wind photoionization
structure and thermodynamics. 

We find that the local disk radiation can launch a wind from the disk
despite  strong ionizing radiation from the central object. The central
radiation may overionize the supersonic portion of the flow
and severely reduce the wind velocity. To produce a fast disk wind
the wind X-ray opacity must be higher than the UV opacity by $\simgreat$~2
orders of magnitude.

Our calculations of a wind from a disk accreting onto a $10^{8}~\MSUN$ black 
hole at the rate of 1.8~$\MSUNYR$ show that the radiation force from 
an accretion disk can launch a self-shielding wind from a radius of 
$\simless 10^{16}$~cm while the strong  UV radiation from the central object 
can radially accelerate the disk wind to velocities 
$\sim 15000$~km~$\rm s^{-1}$, for the X-ray opacity of 40~$\rm g^{-1}~cm^2$. 
The  disk wind domain is intrinsically unsteady and its covering factor
is $\sim 0.2$. The wind mass loss is 0.5~$\MSUNYR$ which 
is a significant fraction of the mass accretion rate.
The strong  X-ray radiation from a central object completely ionizes 
the polar region and only a thin layer above the upper envelope of 
the disk wind. 
The disk wind immediately below the upper envelope can
be characterized as a fast, high-density stream which is reminiscent
of the PSD disk wind solution dominated by the driving radiation from
a bright central object. 
The column density of the fast stream is between 
$10^{22}~\rm cm^{-2}$ and $10^{24}~\rm cm^{-2}$ so the stream 
is optically thin to the UV radiation 
and this is precisely why it can be accelerated to high velocities.  
On the other hand, the part of the wind bounded by
the fast stream, closer to the disk, is slow  and is reminiscent of  
the PSD solution dominated by the driving radiation from the disk. 
The column density of the slow  wind is $>10^{24}~\rm cm^{-2}$ so
it is completely shielded from the central radiation, 
both the X-rays and the UV photons.

ACKNOWLEDGEMENTS: We would like to thank J.E. Drew for useful discussions.
The work presented in this paper was initiated while
DP was a PPARC Research Associate at Imperial College, London  
and it was performed while DP held
a National Research Council Research Associateship at NASA/GSFC.
Computations were supported by NASA grant NRA-97-12-055-154.

\newpage
\section*{ REFERENCES}
 \everypar=
   {\hangafter=1 \hangindent=.5in}

{
  Abbott, D.C. 1982, ApJ, 259, 282

  Arav, N. 1996, ApJ, 465, 617

  Arav, N., Korista, T.K., Barlow, T. A., \& Begelman, M. C. 1995, 
   Nature, 376, 576

  Arav N., \& Begelman, M.C. 1994, ApJ, 434, 479

  Arav N., Li, Z.Y., \& Begelman M.C. 1994, ApJ, 432, 62

  Arav N.,  Shlosman I., \&  Weymann. R. ed. 1997, in ASP Conf. Ser. 128, 
  Mass Ejection from Active Galactic Nuclei, (San Francisco: ASP)

  Barlow, T.A. 1994, PASP, 106, 548

  Balbus, S.A., \& Hawley, J.F. 1998, Rev. Mod. Phys., 70, 1 

  Begelman, M.C., de~Kool, M., \& Sikora, M. 1991, ApJ, 382, 416

  Blandford, R.D., Netzer H., Woltjer L.,  Courvoisier T., \& Mayor M. 1990,
  Active Galactic Nuclei, (Berlin: Springer)

  Blandford, R.D., Payne, D.G. 1982, MNRAS, 199, 883

  Blondin, J.M. 1994, ApJ, 435, 756

  Blondin, J.M., Kallman, T.R., Fryxell, B.A., \& Taam, R.E. 1990, 
  ApJ, 356, 591

  Boroson, T.A., \& Green, R.F. 1994, ApJS, 80, 109 

  Bottorff, M., Korista, K.T., Shlosman, I., \& Blandford, D.R. 1997, ApJ,
  479, 200

  Brandt, W.N., Laor, A., \& Wills, B. J. 2000, ApJ, 528, 637

  Castor, J.I., Abbott, D.C.,  Klein, R.I. 1975, ApJ, 195, 157 (CAK)
  
  Crenshaw, D.M. 1997, in ASP Conf. Ser. 128, Mass Ejection from 
  Active Galactic Nuclei, ed. N. Arav, I. Shlosman, \& R. Weymann 
  (San Francisco: ASP), p. 121

  de Kool, M., \& Begelman, M.C. 1995, APJ, 455, 448 (dKB)

  Drew, J.E., \& Boksenberg, A. 1984, MNRAS, 211, 813 

  Emmering, R.T., Blandford, R.D., \& Shlosman, I. 1992, ApJ, 385, 460  

  Foltz, C.B., Weymann, R.J., Morris, S.L., \& Turnshek, D.A. 1987, 
  ApJ, 317, 450

  Gayley, K.G. 1995, ApJ, 454, 410

  Hamann, F., Barlow, T.A., Cohen, R.D., Junkkarinen, V., \& Burbidge, E.M.
  1997, in ASP Conf. Ser. 128, Mass Ejection from Active Galactic Nuclei, 
   ed. N. Arav, I. Shlosman, \& R. Weymann (San Francisco: ASP), p. 19

  Kaastra, J.S., Mewe, R., Liedahl, D.A., Komossa, S. \& Brinkman, A.C. 2000,
  A\&A, 354, L83

  Kallman, T.R., \& McCray, R. 1982, ApJS, 50 263

  K$\ddot{\rm o}$nigl, A., \& Kartje, J. F. 1994, ApJ, 434, 446

  Korista, K.T., Voit, G.M. Morris, S.L., \& Weymann R.J. 1993, ApJS, 88, 357 

  Krolik, J.H. 1999, Active galactic nuclei: from the central black hole 
  to the galactic environment, Princeton, N.J.: Princeton University Press

  Laor, A., Fiore, F., Elvis, E., \& Wilkes B.J., \& McDowell J.C. 1997, ApJ,
  477, 93

  Murray, N., Chiang, J., Grossman, S.A., \& Voit, G.M. 1995, ApJ, 451, 498 (MCVG)

  Osterbrock, D.E. 1989, Astrophysics of Gaseous Nebulae and Active Galactic
  Nuclei (Mill Valley: University Science Books), chap. 11 
  
  Owocki, S.P., Castor J.I., \& Rybicki, G.B. 1988, ApJ, 335, 914
  
  Pereyra, N.A., Kallman, T.R., \& Blondin, J.M. 1997, ApJ, 477, 368

  Proga, D. 1999, MNRAS, 304, 938

  Proga, D., Stone J.M., \& Drew J.E. 1998, MNRAS, 295, 595 (PSD~I)

  Proga, D., Stone J.M., \& Drew J.E. 1999, MNRAS, 310, 476 (PSD~II)

  Puls, J., Springmann, U., \& Lennon, M. 2000, A\&AS 141, 23

  Scoville, N., \& Norman, C. 1995, ApJ, 451, 510

  Shakura N.I., \& Sunyaev R.A. 1973 A\&A, 24, 337

  Shlosman, I., Vitello, P., \& Shaviv, G. 1985, ApJ, 294, 96

  Stevens, I.R.  1991, ApJ, 379, 310

  Stevens, I.R., \& Kallman T.R. 1990, ApJ, 365, 321

  Stone, J.M., Norman, M.L. 1992, ApJS, 80, 753

  Svensson, R., \& Zdziarski, A.A. 1994, ApJ, 436, 599

  Voit, G.M., Weymann, R.J., \& Korista, K.T. 1993, ApJ, 413, 95

  Weymann, R.J., Morris, S.L., Foltz, C.B., \& Hewett, P.C. 1991, ApJ, 373,
     23

  Weymann, R.J., Scott, J.S., Schiano, A.V.R., \& Christiansen, W.A. 
  1982, ApJ, 262, 497

  Withbroe, G. 1971, in The Menzel Symposium on Solar Physics, Atomic Spectra,
          and Gaseous Nebulae, ed. K.B. Gebbie (NBS Spec. Pub. 353; Washington,
          D.C.: NBS), p. 127

  Zheng, W., Kriss, G.A., Telfer, R.C., Grimes, J.P., \& Davidsen, A.F. 
  1997, ApJ, 475, 469

}

\newpage

\begin{figure}
\begin{picture}(430,350)
\put(0,0){\includegraphics{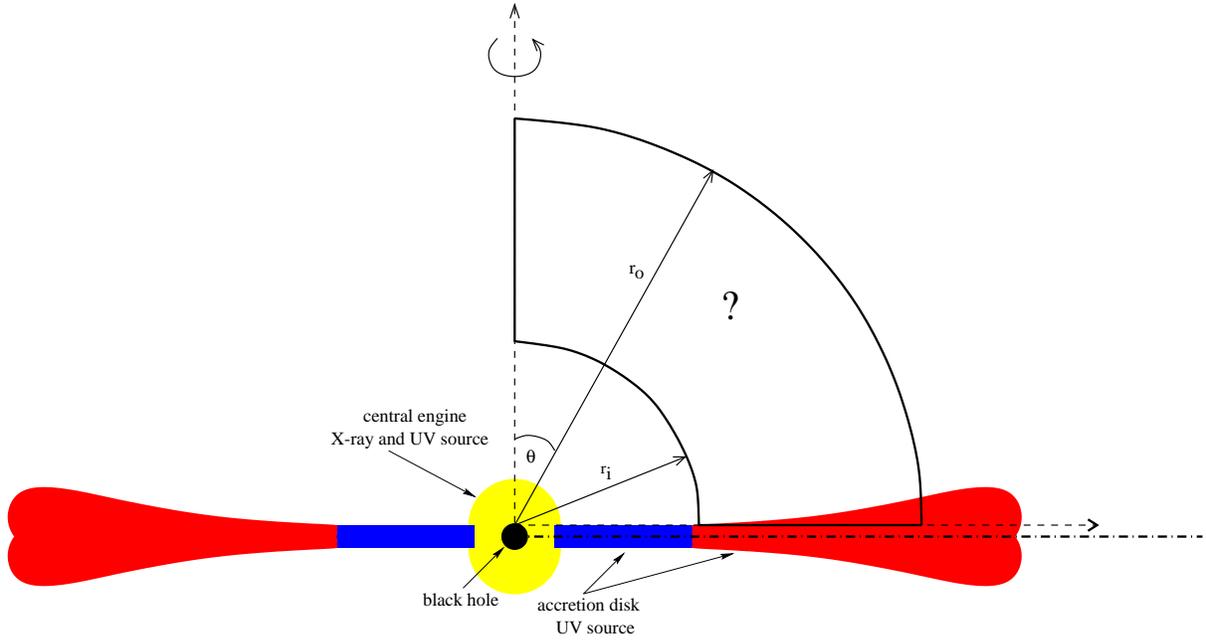}}
\end{picture}
\caption{ The framework of our two-dimensional hydrodynamical calculations
for a line-driven disk wind.  
The drawing is not to scale.
We assume the disk is flat, Keplerian, 
geometrically thin, and optically thick.
The radiation pressure dominated disk is represented by the blue regions while
the gas pressure dominated disk is represented by the red regions.
The black hole is represented by the black circle 
while the X-ray source, is marked by the yellow region. 
The dashed line perpendicular to the disk is the disk rotational axis.
Our computational domain is marked by  solid lines.
The $\theta=~90^o$ axis, the dashed line perpendicular to the rotational
axis, is located above the disk midplane, the dot-dashed line.
The offset of the $\theta=90^o$ axis from the disk midplane is given
by the disk pressure scale height. See Section~2 for more details.
}
\end{figure}

\begin{figure}
\begin{picture}(180,430)
\put(140,0){\includegraphics{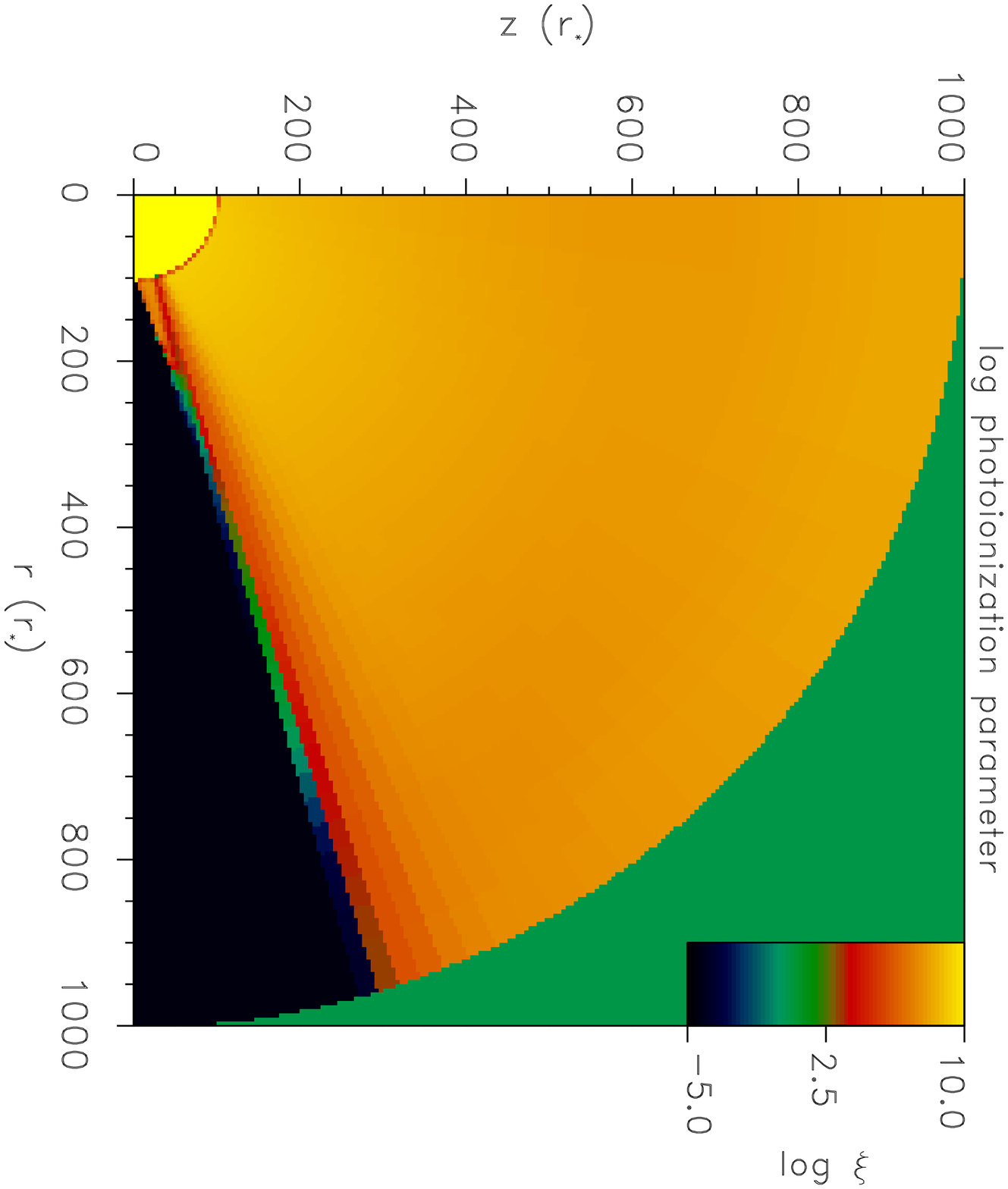}}
\put(140,210){\includegraphics{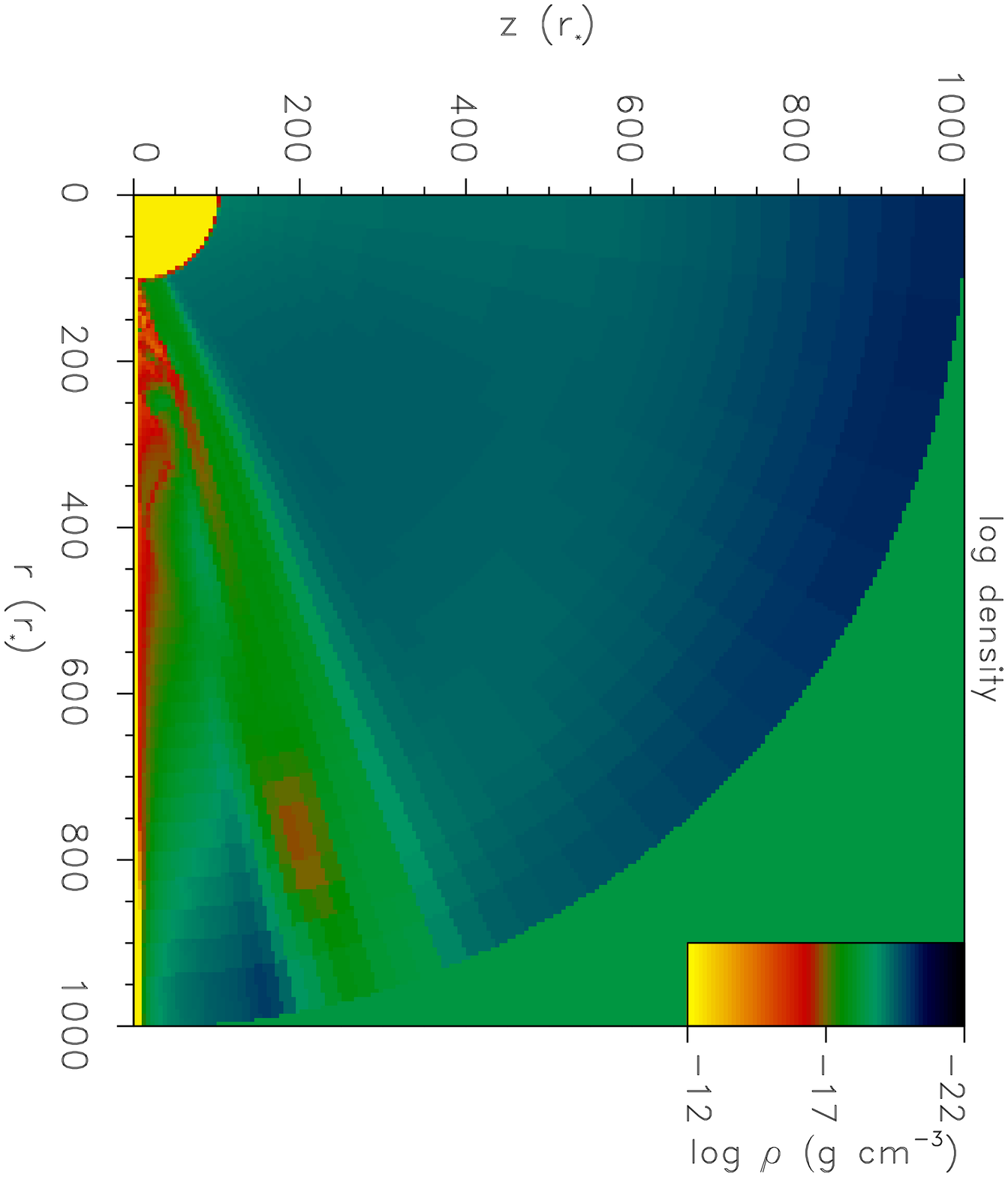}}
\put(230,0){\includegraphics{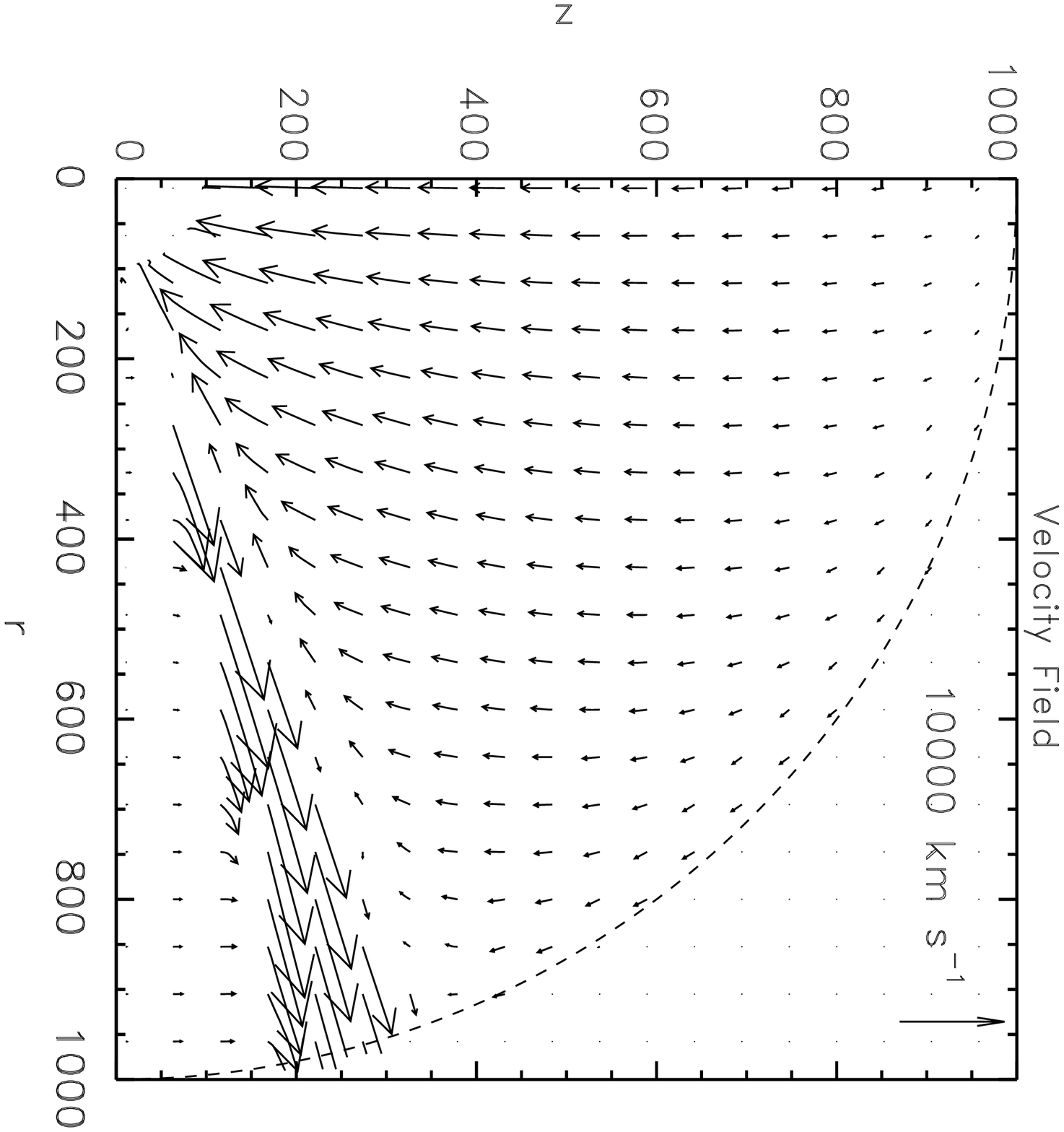}}
\put(230,210){\includegraphics{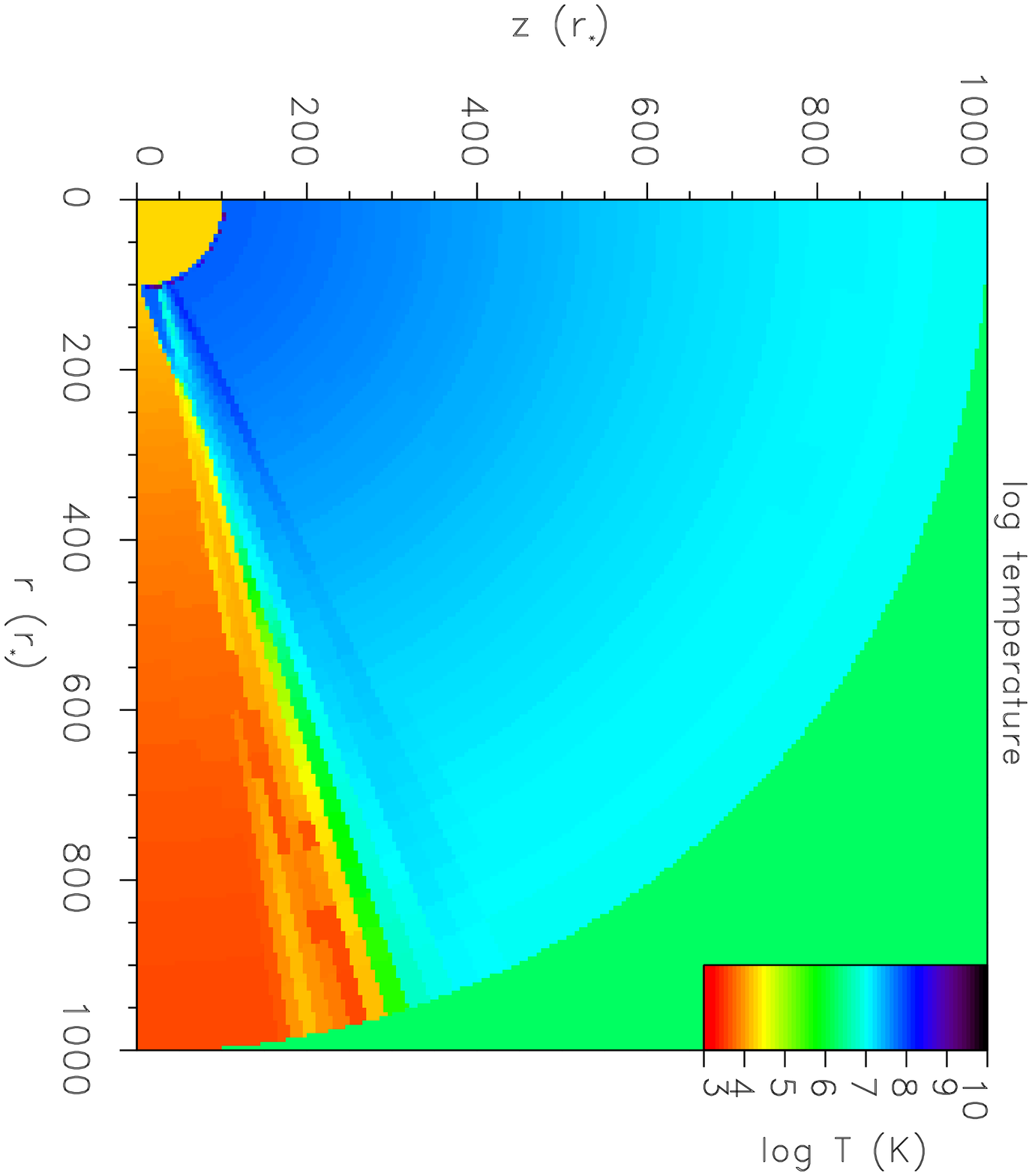}}
\end{picture}
\caption{The top left panel is a color density map of the AGN disk wind
model, described in the text. The top right panel is a color
gas temperature map of the model while the bottom left panel
is a color photoionization parameter map. Finally, the bottom right
panel is a map of the velocity field (the poloidal component only). 
In all panels the rotation axis of the disk is along the left hand vertical 
frame, while the midplane of the disk is along the lower horizontal frame.
}
\end{figure}

\begin{figure}
\begin{picture}(180,230)
\put(129.5,10){\includegraphics{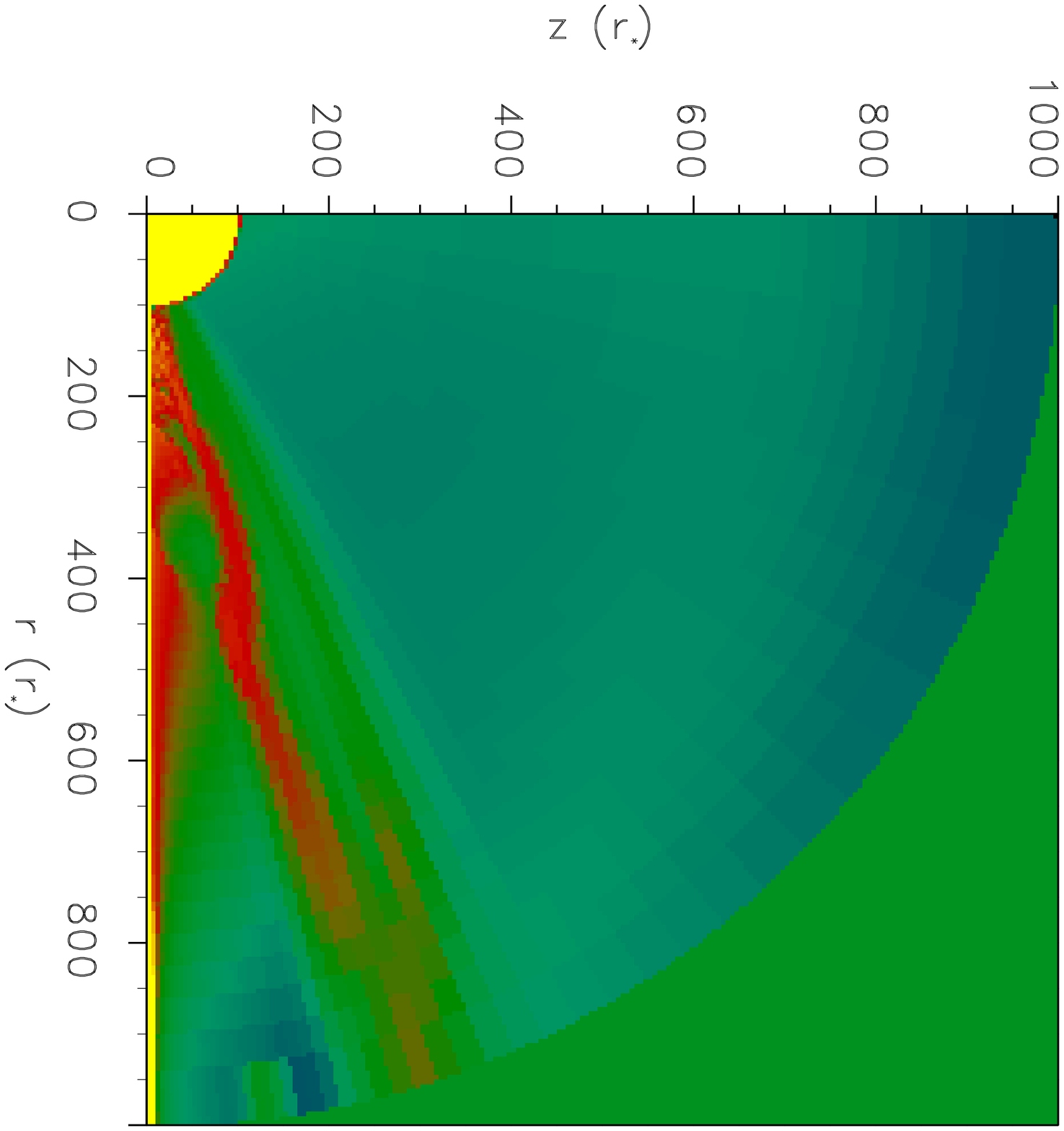}}
\put(130,10){\includegraphics{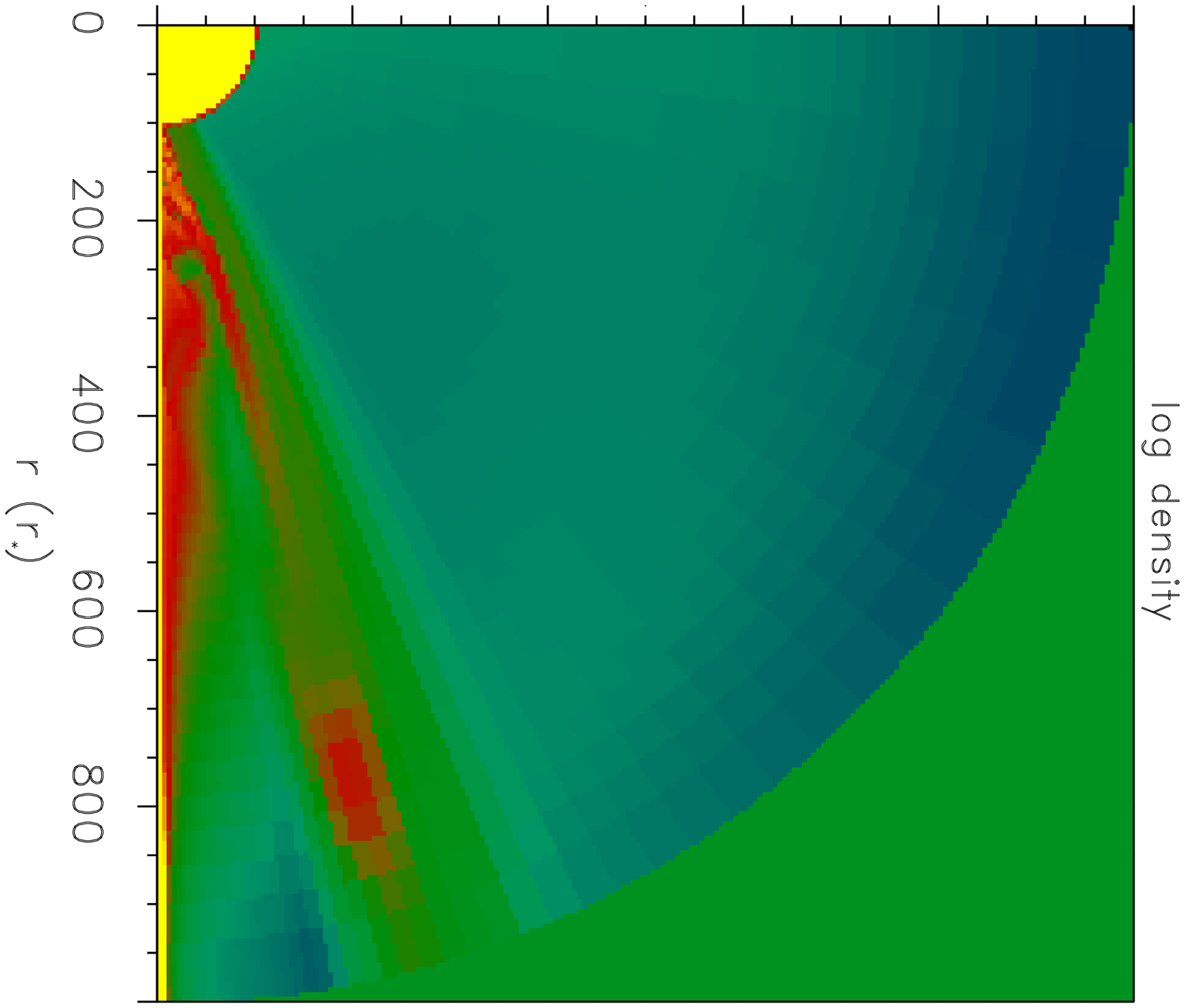}}
\put(286,10){\includegraphics{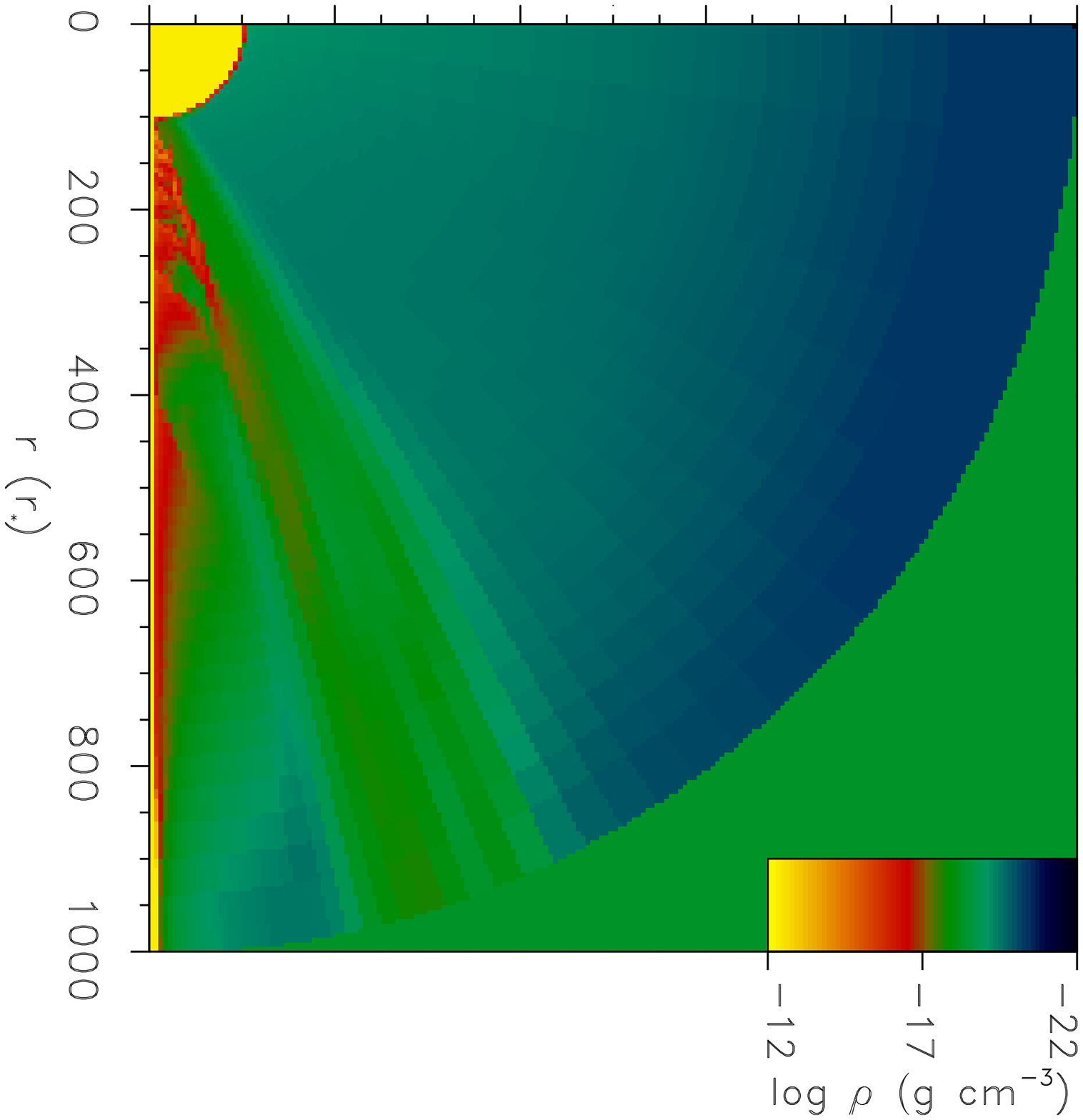}}
\end{picture}
\caption{A sequence of density maps for the model from Figure~2 after
13.3, 14.6, and 16.47~years (left, middle, and right panel).
Note both the time-dependent fine structure near the base of the wind
and the time-dependent large scale structure
associated with the fast stream at the polar angle $\sim 75^o$}.
\end{figure}

\begin{figure}
\begin{picture}(600,400)
\put(0,0){\includegraphics{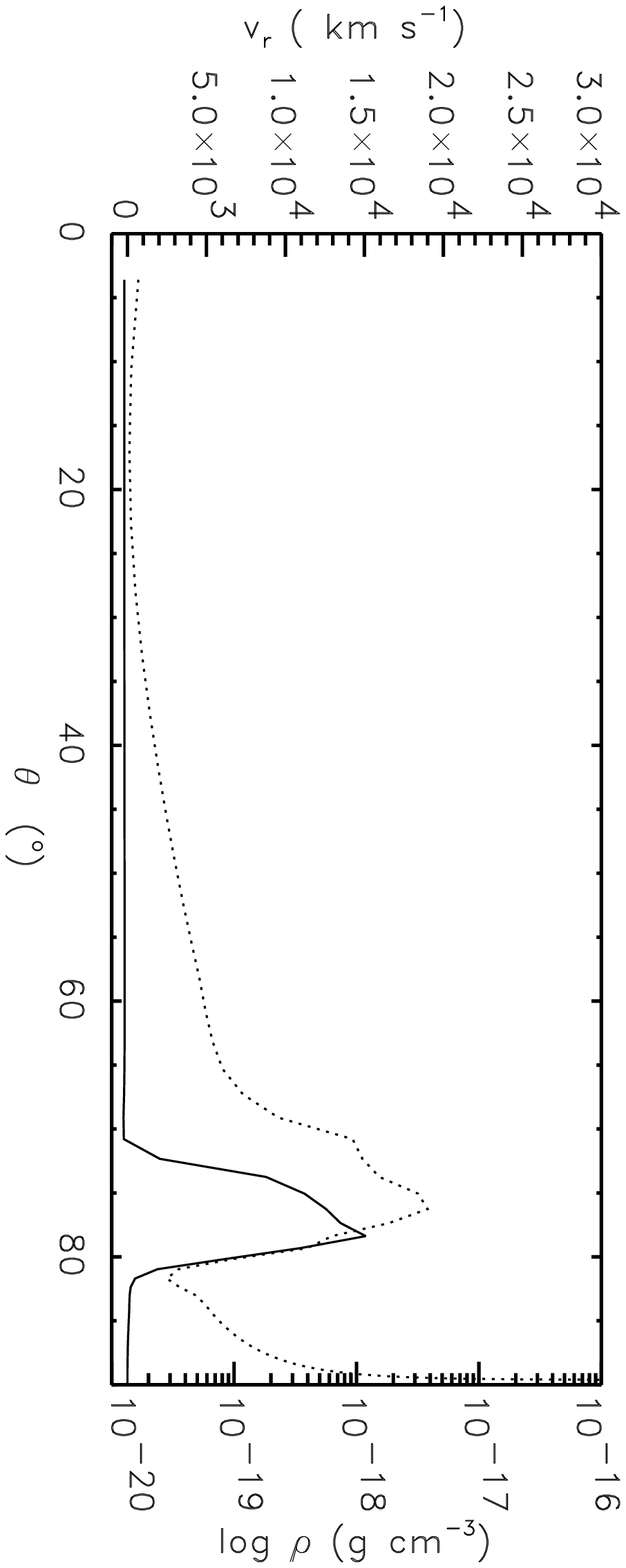}}

\put(0,0){\includegraphics{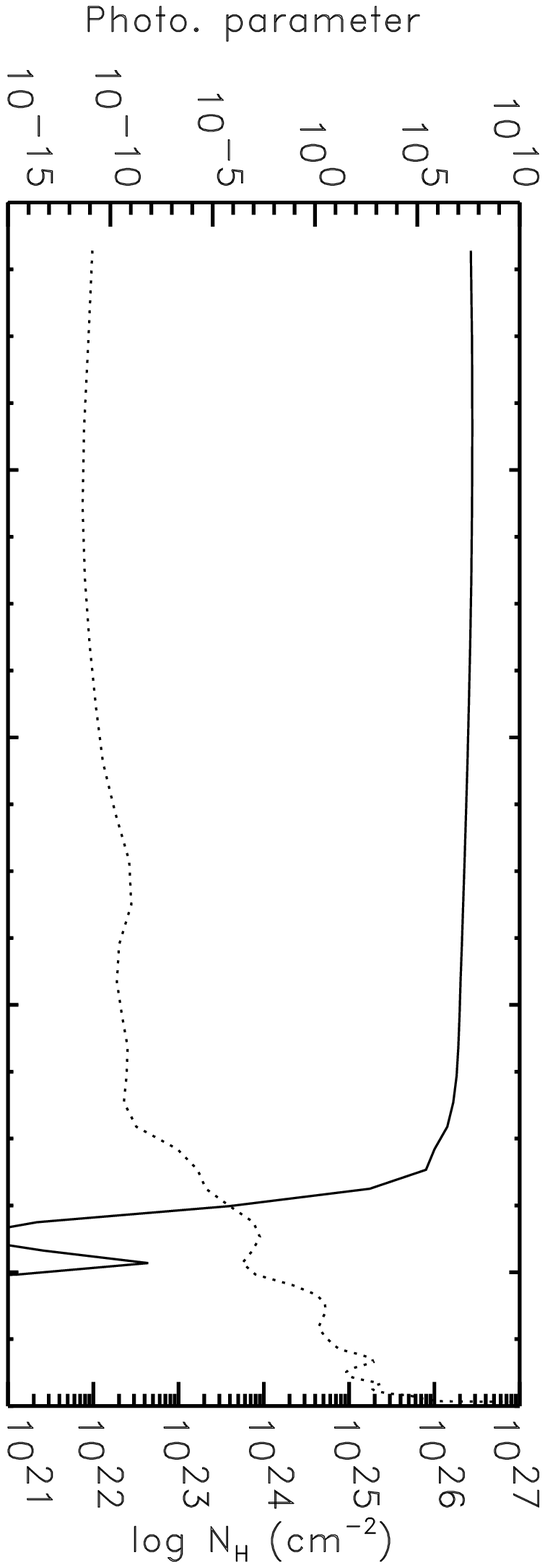}}
\end{picture}
\caption{Quantities at the outer boundary in our model (the top left
panel of Figure~2 and the middle panel of Figure~3).
The ordinate on the left hand side of each panel refers to the solid
line, while the ordinate on the right hand side refers to the dotted
line. 
The column density, $\rm N_H$ is calculated  along the radial direction.
The radial velocity peaks at $\theta \approx 74^o$
while the density at $\theta \approx 71^o$. The column
density is less than $10^{24}~\rm cm^{-2}$ for $\theta \simless 80^o$.}
\end{figure}

\end{document}